\documentclass{emulateapj}
\usepackage{graphicx,natbib,apjfonts}
\usepackage{amsmath}
\usepackage{amssymb}
\usepackage{epstopdf}
\usepackage{xfrac}
\usepackage{rotate}
\begin{document}

\date{\today}
\title{THE CLUSTERING OF GALAXIES AROUND RADIO-LOUD AGN{s}}
\author{
Hauke Worpel\altaffilmark{1},
Michael J. I. Brown\altaffilmark{2},
D. Heath Jones\altaffilmark{2},
David J. E. Floyd\altaffilmark{2},
Florian Beutler\altaffilmark{3}\altaffilmark{4}
}
\altaffiltext{1}{Monash Centre for Astrophysics, School of Mathematical Sciences, Monash University, Clayton, Victoria 3800, Australia}
\altaffiltext{2}{Monash Centre for Astrophysics, School of Physics, Monash University, Clayton, Victoria 3800, Australia}
\altaffiltext{3}{ICRAR, University of Western Australia, 35 Stirling Highway, Perth WA 6009, Australia}
\altaffiltext{4}{Lawrence Berkeley National Laboratory, 1 Cyclotron Road, Berkeley, CA 94720, USA}
\slugcomment{ApJ (accepted)}
\begin{abstract}
We examine the hypothesis that mergers and close encounters between galaxies can fuel AGNs by increasing the rate at which gas accretes towards the central black hole. We compare the clustering of galaxies around radio-loud AGNs with the clustering around a population of radio-quiet galaxies with similar masses, colors and luminosities. Our catalog contains 2178 elliptical radio galaxies with flux densities greater than 2.8~mJy at 1.4~GHz from the 6dFGS survey. We find that radio AGNs with more than 200 times the median radio power have, on average, more close (r<160~kpc) companions than their radio-quiet counterparts, suggestive that mergers play a role in forming the most powerful radio galaxies. For ellipticals of fixed stellar mass, the radio power is not a function of large-scale environment nor halo mass, consistent with the radio powers of ellipticals varying by orders of magnitude over billions of years.

\end{abstract}

\maketitle

\section{Introduction}

Radio-loud active galactic nuclei (RLAGNs) predominantly inhabit massive elliptical galaxies (e.g., \citealt{DunlopEtAl2003, FloydEtAl2004}) and are powered by the infall of material onto a supermassive black hole (SMBH) at the core of the host galaxy \citep{Salpeter1964, LyndenBell1969}. It is not yet clear what reservoir of gas the RLAGN uses~\citep{BaumHeckman1989, McCarthy1993, CoilEtAl2009}. Gas can be transported into the inner regions of the galaxy through angular momentum losses caused by mergers with other galaxies (e.g.~\citealp{BarnesHernquist1991, BarnesHernquist1996}), or through cooling and accreting stochastically~\citep{TaborBinney1993, HopkinsHernquist2006}. This paper seeks to distinguish between these mechanisms.

It is possible to find evidence for these mechanisms by studying the environments of the AGN host galaxies. If the emission is driven by mergers or close encounters with other galaxies then we should expect to see an excess of close companions to RLAGN hosts (compared to massive ellipticals that do not show RLAGN activity). Under one hypothesis, weak radio emission may be powered by gentle, stochastic accretion while strong emission requires violent events such as mergers and close encounters to supply a large amount of gas all at once~(e.g., \citealt{ShlosmanEtAl1990, HopkinsHernquist2006}). If this "mixed" hypothesis is true, then we expect the strength of clustering around AGN hosts to increase as the radio power increases. 

Observations increasingly point to multiple modes being responsible for the triggering of AGNs. A general picture has emerged in which radio-loud AGNs ($\gtrsim~10^{23}$~W~Hz$^{-1}$) occupy richer environments than low power ($\lesssim~10^{23}$~W~Hz$^{-1}$) radio sources, with a general increase in clustering strength with radio luminosity (e.g., \citealt{CroftEtAl2007, WakeEtAl2008}). The most luminous sources ($\gtrsim~10^{25.5}$~W~Hz$^{-1}$) are more prone to morphological irregularities than less powerful ones (e.g.~\citealt{HeckmanEtAl1986}), indicative of interactions and ongoing mergers with close companions. In contrast, \cite{MagliocchettiEtAl2004} find that, while radio AGNs are more clustered than radio-quiet galaxies, there is no evidence that the strength of this clustering increases over the range $10^{20}-10^{25}$~W~Hz$^{-1}$.

Strong radio sources are known to generally inhabit dense environments (e.g., \citealt{BahcallEtAl1969, LongairSeldner1979, PrestagePeacock1988, Best2000}), which is consistent with the merger hypothesis. However, this alone is not a telltale sign of the merger hypothesis because both dark matter halo mass and radio luminosity of an RLAGN host increase with stellar mass (e.g., \citealt{LiEtAl2006b, MosterEtAl2010}), which would also result in RLAGNs occupying dense environments.

Much observational work in this area has investigated the statistical properties of large numbers of galaxies, making use of the two-point cross-correlation function to compare the clustering of galaxies around RLAGN hosts with that around radio-quiet galaxies that are otherwise physically similar (e.g., \citealt{YeeGreen1987, SadlerEtAl2007, WakeEtAl2008}). In this way, any correlation of host stellar mass with halo mass and radio luminosity can be identified.

We have compared the clustering properties of radio-loud active galactic nuclei with those of a control population selected with the same color, luminosity, and morphology, using the two-point cross-correlation function. In \S \ref{sec:Galaxycatalogs} we describe our data, details of its collection, its allocation into various catalogs, and the generation of random galaxy catalogs. Section \ref{sec:CrossCor} reviews the theory of the two-point cross-correlation function and its application to the galaxy and mock catalogs. In \S \ref{sec:Results} we present our results, and in sections \S \ref{sec:Discussion} and \S \ref{sec:Conclusions} we discuss their meaning and draw conclusions from them. Errors are given as the $1\sigma$ significance and are assumed to be normal, unless otherwise stated. We adopt a cosmology with $H_0~=~72$~km~s$^{-1}$~Mpc$^{-1}$, $\Omega_m=0.25$, $\Omega_\Lambda=0.75$, and $\Omega_K=0$.

\section{Galaxy catalogs}
\label{sec:Galaxycatalogs}

Our goals are to determine whether RLAGNs are more likely to have nearby companion galaxies than a physically similar population of controls that do not show any AGN activity, and whether there is any dependence on large scale environment. We begin with a \emph{main sample} catalog of galaxies and identify which are our RLAGNs of interest. Then we select from the same catalog a number of controls that match the RLAGNs in color and luminosity but which are not radio-loud, and count the number of main sample galaxies within a range of radii of the control or RLAGN galaxy in question. These counts are compared with a prediction of how many companions the central galaxy ought to have if there were no clustering effects, using a \emph{random catalog} representing a totally homogeneous population.

\subsection{6dFGS Main Sample}
\label{sec:NormalGalaxies}

The initial galaxy sample is photometrically selected from the 2MASS Extended Source Catalog (2MASS XSC;~\citealt{JarrettEtAl2000}). This survey includes a complete census, covering most of the sky, of extended objects with total K-band magnitudes of $K<13.5$~\citep{JarrettEtAl2000b}. Redshift data is taken from the 6 Degree Field Galaxy Survey (6dFGS), a combined redshift and peculiar velocity survey which covers most of the southern sky, excluding ten degrees above and below the Galactic plane~\citep{JonesEtAl2004, JonesEtAl2005, JonesEtAl2009}. Spectra were taken with the Six-Degree Field (6dF) multifiber spectrograph, an instrument able to simultaneously collect up to 150 different spectra over the 5.7 degree field of view of the UK Schmidt telescope at Siding Spring.

For the purpose of this paper, the targets of the 6dFGS survey are all galaxies listed as being brighter than magnitude 12.75 in the $K$-band in the original 6dFGS target catalogue \citep{JonesEtAl2004}. The final redshift completeness of the survey is around 90\% \citep{JonesEtAl2009}. Redshifts were calculated using the cross-correlation technique of \cite{TonryDavis1979} and the quality checked by visual inspection. The availability of spectroscopic (rather than photometric) redshifts allows galaxy distances to be determined with high accuracy and precision. Typical errors on the redshift measurements are $\Delta cz \approx 50$~km~s$^{-1}$ \citep{JonesEtAl2009}. 

Subsequent refinements of the 2MASS photometry led to the inclusion and exclusion of galaxies near the 12.75 K-band limit \citep{JonesEtAl2009} so we follow the same procedure as \cite{BeutlerEtAl2011} and impose $K \leq 12.9$, leaving 81,971 galaxies. Although our radio and control galaxies (see \S~\ref{sec:RadioGalaxies} and \S~\ref{sec:ControlGalaxies}) are restricted to objects with declinations greater than $-40^\circ$, we do not use the same restriction on the main sample. This is done to increase the number of 6dFGS main sample galaxies and maximize the signal.

\subsection{Radio Galaxies}
\label{sec:RadioGalaxies}

The radio galaxy catalog is generated by combining data from 6dFGS (DR3; \citealt{JonesEtAl2009}) and the 1.4~GHz NRAO VLA Sky Survey \citep{CondonEtAl1998}. The objects in 6dFGS identified as radio galaxies are from the 1.4 GHz NRAO VLA Sky Survey (NVSS), a catalog of $\sim2\times10^6$ discrete radio sources with flux densities greater than $S\approx 2.5$ mJy. The images have a FWHM of $\theta \approx 45''$~\citep{CondonEtAl1998}. A list of objects in the second data release of the 6dFGS database matching radio sources in the NVSS was produced by Mauch and Sadler (2007)\nocite{MauchSadler2007}. It contains 7,824 objects with K-band magnitudes brighter than 12.75, flux densities greater than 2.8~mJy at 1.4~GHz and declinations between -40$^\circ$ and $0^\circ$. \cite{MauchSadler2007} distinguished AGN hosts from star-forming galaxies by visually inspecting their spectra. Mauch later updated the list with the third 6dFGS data release, increasing it to 9,296 objects; we use this updated list (Mauch, T.; private comm). 

$B_J$ and $R_F$-band magnitudes are taken from SuperCOSMOS scans of UK Schmidt telescope plates \citep{HamblyEtAl2001}. We have restricted our sample to radio sources with good 6dFGS optical spectra that were classified as AGNs by \cite{MauchSadler2007} and that also have $B_J$ and $R_F$-band photometry.

To remove blue galaxies we impose an empirical color cut: galaxies with $B_J-R_F$ colors less than $0.9+4z$ were removed from the sample. This color cut is shown in Figure \ref{fig:Rvz}, and selects an early-type sample that incorporates the necessary $k$-correction. When the $k$-correction formulae from \cite{ColeEtAl2005} are applied to our color cut (in our own rest frame) the result for $0.02<z<0.25$ is a horizontal line corresponding to a color cut of $B_J-R_F=1.15$ in the rest frame of the galaxies (to within 0.1 magnitude), so that the galaxies retained are redder than $B_J-R_F\approx 1.15$ in their own rest frame. Therefore we deem that the uncorrected color cut is sufficient. We have also excluded galaxies with a spectroscopic redshift of less than $z=0.02$ because we want galaxies with negligible peculiar motions relative to the velocity of the Hubble flow. At $z=0.02$ the recession velocity is about 6,000~km/s so a velocity dispersion inside a galaxy cluster of 500~km/s (e.g., \citealt{GirardiEtAl1993}) is less than a 10\% effect.

To minimize remaining contamination by star-forming galaxies, we excluded disk galaxies from the radio galaxy catalog by cross-matching with morphological information from the HyperLeda catalog \citep{PaturelEtAl2003}. There are 57,451 6dFGS galaxies between -40$^\circ$ and $0^\circ$ in declination. Of these, 55,848 have morphological information in HyperLeda. Any galaxy identified as a spiral was removed. Finally we discarded galaxies with a minor to major axis ratio of 0.4 or less in either the K-band 3$\sigma$ isophote or the stacked J+H+K coadded $3\sigma$ isophote using the 2MASS database. These likely contain significant numbers of edge-on spirals, many of which can make our color cut due to reddening caused by dust lanes. Excluding disk galaxies is very important because these may have weaker clustering properties than spheroidal galaxies, and so spiral contamination of the control galaxies could enhance the difference between the radio galaxy environments and the control galaxy environments. Despite these precautions it is likely that a small level of disk galaxy contamination still remains. We visually inspected SDSS images of the first hundred radio galaxies retained, sorted by decreasing declination because the SDSS coverage is more complete near the equator. There was one contaminant: a face-on spiral galaxy with very red color, meaning that the contamination by disk galaxies is unlikely to exceed 5\%. After removing these contaminants, 2,126 galaxies remained.

We found by visual inspection of SDSS images of NVSS sources that a number of red ellipticals are missing from the \cite{MauchSadler2007} catalog. This is because Mauch's catalogue did not include 6dFGS sample galaxies that had spectra/redshifts sourced from other surveys. The majority of these have Programme ID 1 in the 6dFGS database, meaning that they belong to the Main Target sample (K-band selected; see \citealt{JonesEtAl2009}). Since these objects are missing Mauch's spectral classifications, we only add objects that have good SDSS imaging where star-forming or disk contaminants are easily identified using the SDSS imaging. There are 52 such galaxies, bringing the total up to 2,178.

A galaxy with a given radio luminosity may be a large galaxy emitting modestly, or a small galaxy emitting very strongly for its size. Any analysis based upon radio luminosity cuts alone will treat both the same way. The importance of taking this effect into account has been pointed out previously (e.g., \citealt{HickoxEtAl2009, MandelbaumEtAl2009, DonosoEtAl2010}), but no previous study has explicitly studied the clustering strength of RLAGNs with increasing \emph{relative} radio luminosity. We compare the observed radio luminosity of each galaxy to the median radio luminosity of galaxies of the same K-band luminosity. \citet{BrownEtAl2011} find that this median is well-fit by
\begin{equation}
\mathfrak{L}(L_K)=1.16\times10^{20}\text{W Hz}^{-1}\times\left(\dfrac{L_K}{10^{11}L_{\odot}}\right)^{2.78}\times1.4\text{GHz},
\label{eqn:Upsilon-definition}
\end{equation}
where ${L_K}$ is the K-band luminosity of the galaxy in solar luminosities. 

We introduce $\Upsilon$, the observed radio luminosity of an individual radio galaxy divided by its predicted median radio luminosity as given by equation \ref{eqn:Upsilon-definition}:
\begin{equation}
\label{eqn:PhiDefinition}
\Upsilon\equiv\dfrac{L_{\text{radio,obs}}}{\mathfrak{L}(L_K)},
\end{equation}
where $L_{\text{radio,obs}}$ is the observed radio luminosity of the galaxy.

To investigate the effect of increasing radio luminosity we divide the radio galaxies into $\Upsilon$ bins with boundaries at $\Upsilon=$10, 20, 40, 80, and 200 populated as shown in Table \ref{tab:CompanionCounts}.

\subsection{Control Galaxies}
\label{sec:ControlGalaxies}

Since radio AGNs typically inhabit massive ellipticals, and massive ellipticals are preferentially found near the centers of their dark matter haloes, we expect enhanced clustering for them relative to less massive galaxies, regardless of radio AGN activity. To distinguish any additional effect from this natural clustering we must compare the radio AGN hosts with a population of galaxies with similar physical properties but which are not radio AGNs.

We restrict our radio-quiet galaxy set to 6dFGS galaxies with declinations $-40^\circ$ to $0^\circ$ and redshifts $z>0.02$, with the same color cut as the radio galaxies. 24,240 objects remained after excluding disk galaxies identified in the same way as in the RLAGN sample, objects in the complete catalog of \cite{MauchSadler2007}, and objects within 12" of a radio source in NVSS. Since we used the spectral classifications of \cite{MauchSadler2007} to exclude star-forming galaxies there may be slightly more contaminants in the control sample. This is because no similar classification of radio-quiet galaxies in 6dFGS exists.

\begin{figure*}
\begin{center}
\includegraphics[width=84mm]{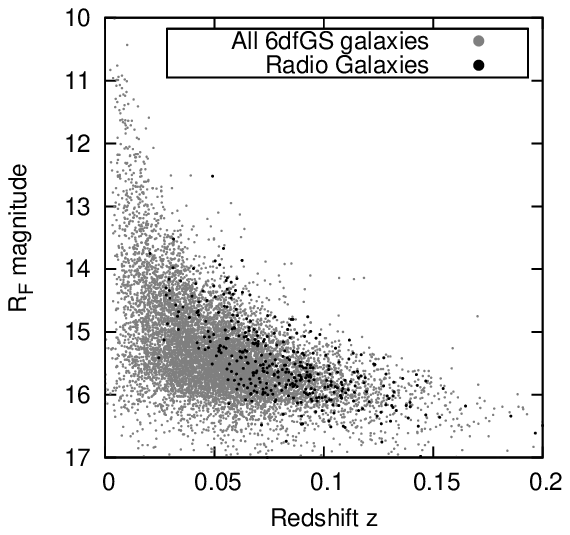}\includegraphics[width=84mm]{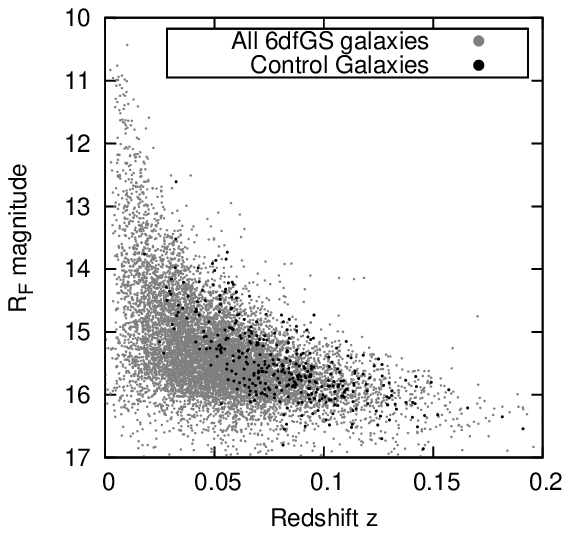}\\
\includegraphics[width=84mm]{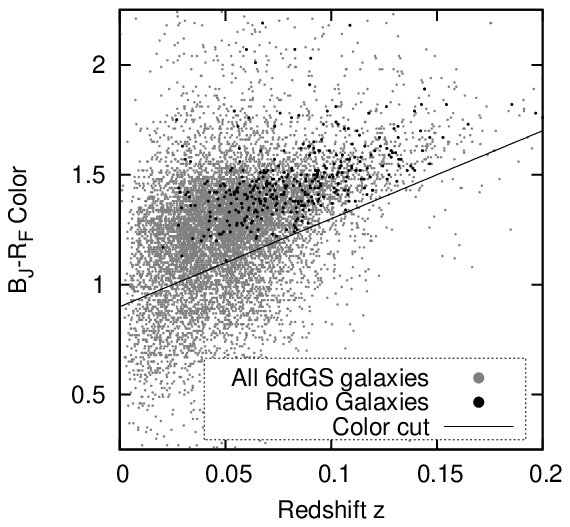}\includegraphics[width=84mm]{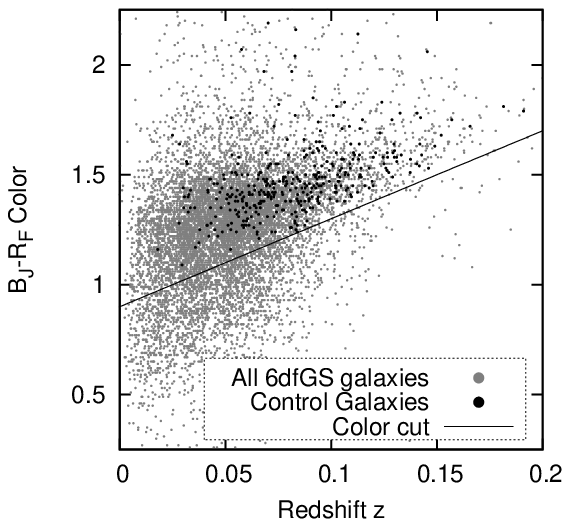}\\
\end{center}
\caption{Redshift distributions of \emph{z}~>~0.02 radio galaxies (left panels) and controls (right panels) in $R_F$-magnitude and $B_J-R_F$ color space. Shown here are the radio and control galaxies corresponding to the $40<\Upsilon<80$ subset. The control galaxies are chosen from the restricted 6dFGS data to lie as close as possible to the radio galaxies in these spaces. Every radio galaxy, every fourth control galaxy, and every tenth main sample galaxy is plotted. Also shown is the color cut applied to the radio and control galaxies ${B_J-R_F}>0.9+4z$.}
\label{fig:Rvz}
\end{figure*}

For each radio galaxy we selected four radio-quiet objects from the restricted 6dFGS data to act as controls. In order to approximately match the color and redshift distribution of the radio galaxies we picked the four nearest objects in four-dimensional ($B_J$-magnitude, $R_F$-magnitude, redshift, ($B_J-R_F$) color) space such that the quantity
\begin{equation}
D^2\!=\!\!{ \left(\dfrac{\Delta B_J}{10.5}\right)^2\!\!\!+\!\left(\dfrac{\Delta R_F}{10.5}\right)^2\!\!\!+\!\left(\dfrac{\Delta z}{0.3135}\right)^2\!\!\!+\!\left(\dfrac{\Delta(B_J-R_F)}{1.25}\right)^2 }
\end{equation}
was minimized (see Figure \ref{fig:Rvz}), where $\Delta B_J$, $\Delta R_F$, $\Delta z$, and $\Delta(B_J-R_F)$ are the differences in $B_J$-band magnitude, $R_F$-band magnitude, redshift and $B_J-R_F$ color between the radio galaxy and candidate control galaxy respectively. The scaling factors are essentially arbitrary; we based them on the ranges spanned by the quantities in question, and this gave good results: 90\% of controls were within 0.006 in redshift, 0.06 in B-band magnitude, 0.06 in R-band magnitude, and 0.09 in $B_J-R_F$ color of their radio galaxies.

There is some ambiguity about what to do with radio galaxies with $\Upsilon$ less than the current $\Upsilon$ bin. These could be excluded altogether, or returned to the pool of potential controls. If RLAGNs do have stronger clustering properties than the controls, treating low power radio galaxies as potential controls could reduce the measured difference in clustering. We tested both methods and found that the measured $r_0$ and $\gamma$ values (see \S \ref{sec:CrossCor}) changed by less than 2.5\%. We therefore chose to exclude radio galaxies from the pool of potential controls.

\subsection{Random galaxies}
\label{sec:RandomGalaxies}

\begin{figure}
\includegraphics[width=80mm]{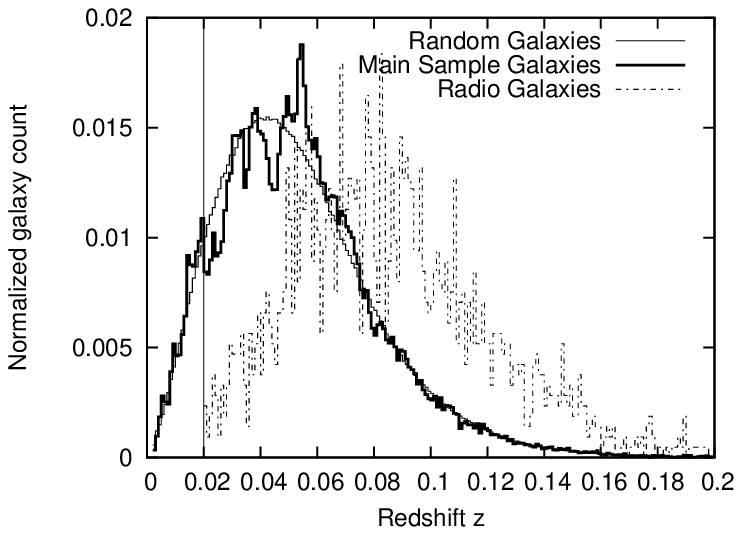}
\caption{The redshift distributions of the 6dFGS main sample (thick line), the random galaxy catalog (thin line) and the radio galaxies (dashed line). The counts have been normalized to unity. The random galaxy and 6dFGS main sample distributions peak at a redshift of around $z=0.04$ while the radio galaxies peak at a redshift of around $z=0.07$. The $z=0.02$ redshift bin applied to the 6dFGS and random galaxies is indicated with a vertical black line.}
\label{fig:MainRandomRedshifts}
\end{figure}

A good random catalog should simulate what observers would see if they were looking at a homogeneous distribution of galaxies. It should reproduce all observational incompleteness, but it should not reproduce features in the observed data that are due to clustering since this is the property we intend to measure. Our random catalog is based upon the 6dFGS luminosity function of \cite{JonesEtAl2006} and is generated according to the procedure outlined in~\cite{BeutlerEtAl2011}, using the finalized 6dFG survey mask. It contains 2,459,130 objects: thirty for every main sample galaxy. The redshift distributions of the random catalog and main sample catalogs are shown in Figure \ref{fig:MainRandomRedshifts}. A large random catalog minimizes noise, and allows the correlation function to be measured at very small and very large separations where there are few data-data pairs. The random sample covers the same area as 6dFGS, including regions without NVSS coverage.

\section{The Two-Point Cross-Correlation Function}
\label{sec:CrossCor}

We will apply the two-point cross-correlation function to the galaxy and mock catalogs described in \S~\ref{sec:Galaxycatalogs}. We calculate the cross-correlation function $\xi(\sigma,\pi)$ using pair counts allocated to a two-dimensional grid of logarithmically spaced bins 0.20 dex in size, beginning with separations of 10~kpc in the radial direction and 10~kpc in the transverse direction. Following \cite{MeyerEtAl2007} we ignore pairs with an angular separation of more than 50 degrees. At $z=0.02$ (83~Mpc) an angular separation of 50 degrees corresponds to a physical separation of 70~Mpc, much larger than the scales we investigate. Furthermore, wide-angle effects have been found to have a negligible effect on 6dFGS clustering measurements \citep{BeutlerEtAl2011, BeutlerEtAl2012}. We do not use any additional magnitude restrictions on the pair counts.

The fiber buttons used in the 6dFGS have a radius of 5.6 arcmin on the sky, and galaxy pairs at separations of less than 1~Mpc may be under-represented due to limitations in fiber proximity by up to 20\% \citep{CampbellSaundersColless2004}. Although the vast majority of the 6dFGS targets were 2MASS selected, and therefore not preferentially radio galaxies, there were a small number of NVSS objects assigned low priority. Since such galaxies were only observed when no fibers could be assigned to the high priority targets, this has the potential to bias our analysis towards radio galaxies in crowded fields. However, all radio galaxies in our sample are flagged with Programme ID 1 in the 6dFGS database, meaning that none of the low-priority NVSS galaxies are included in our radio sample, and so there is no 6dFGS target bias within our sample \citep{JonesEtAl2009}. We can therefore be very confident that, although the number of companions counted may be slightly lower than the true value, the problem of fiber collisions will affect radio galaxies and controls equally.

At low angular separations there are also few pairs in any of the bins, leading to noise dominating the signal. This is unfortunate because it is the clustering properties at low separations that we are particularly interested in. We cannot do anything about the real galaxies, but we can reduce the amount of noise arising from the randoms. At low angular separations the $\sigma$ bins are small compared to the scale of irregularities in the 6dFGS on-sky completeness function. We can therefore replace a random scattering of galaxies with a homogeneous distribution whose density is calculated from larger and more populated bins. This is done as follows: for every $\pi$ bin we identify the smallest $\sigma$ bin with 25 or more data-random pairs. If this is less than 6~Mpc in size we calculate the mean pair density of that bin and the next two larger ones, and use this mean density for all the smaller $\sigma$ bins with 9 or less pairs in the original count. The results did not change appreciably if we smoothed using four or five bins rather than three, or set the lower pair limit to 50 rather than 25. We are therefore confident of the robustness of the method. The distance at which we switch to the smoothing method is typically 50-200~kpc.

From $\xi(\sigma, \pi)$ we calculate the \emph{projected correlation function} $\Xi(\sigma)$, which is the integral of $\xi(\sigma, \pi)$ along the line of sight:
\begin{equation}
\Xi(\sigma)=2\int^{D_\text{lim}}_0\xi(\sigma,\pi)d\pi.
\label{eqn:Xi-integral}
\end{equation}
The purpose is to mitigate radial distortions of the galaxy distributions in redshift space caused by motions of galaxies in space superimposed upon the cosmological redshift. These distortions include the "Fingers of God" effect, which causes galaxy clusters to appear elongated, and large-scale flattening caused by the coherent motion of galaxies towards large regions of enhanced density~\citep{Kaiser1987}. The limit $D_\text{lim}$ is imposed because the integral is generally not convergent~\citep{MeyerEtAl2007}; at large radii the correlation function is dominated by large random noise associated with the low pair counts. In this work we adopt $D_\text{lim}=33$~Mpc, the size of one of our radial bins. Changing $D_\text{lim}$ to include one more or one fewer bin (20.8 or 52.3~Mpc) did not significantly affect the results.

Errors on $\Xi(\sigma)$ are estimated using jack-knife resampling. The data samples are divided into eighteen right ascension bins, each containing approximately equal numbers of main-sample galaxies, and $\Xi(\sigma)$ is recalculated, each time omitting one of the RA bins. If we assume a power law form for $\xi(r)$, then the integral in equation \ref{eqn:Xi-integral} can be solved in closed form, and $r_0$ and $\gamma$ directly calculated ~\citep{DavisPeebles1983} using a least-squares linear fit to the logarithms of $\Xi(\sigma)$ and $\sigma$ for the whole sample. The uncertainties were found by repeating the power law fits for each of the jack-knife subsamples.

To compare the clustering around radio galaxies with the clustering around radio-quiet control galaxies (of similar physical properties) we calculate correlation functions for both. The behaviour of clustering as a function of radio galaxy luminosity is investigated by dividing the radio galaxy catalog into $\Upsilon$ bins, generating control galaxy catalogs for each $\Upsilon$ bin, and calculating clustering measures again. We also repeat the procedure for a variety of radio luminosity bins so that our results may be placed in context with previous literature, though this may be a less sensitive selection criterion than one that takes into account the mass of the target galaxies.

\section{Results}
\label{sec:Results}

\begin{figure*}[h!]
\includegraphics[width=80mm]{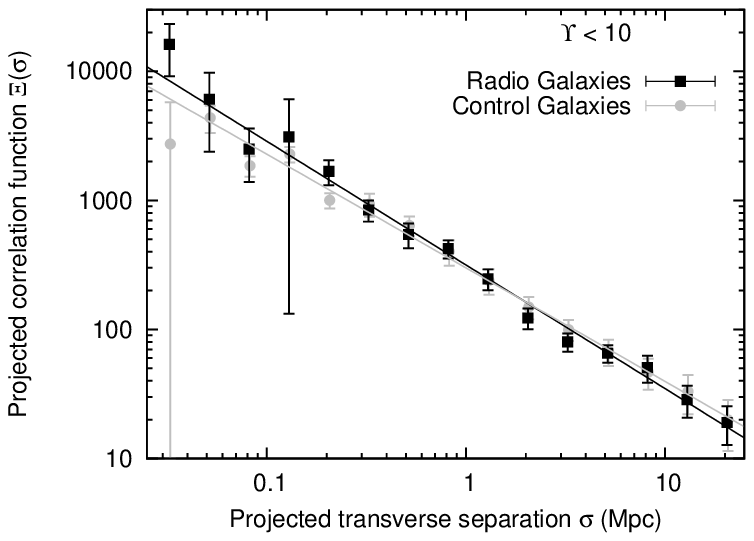}\includegraphics[width=80mm]{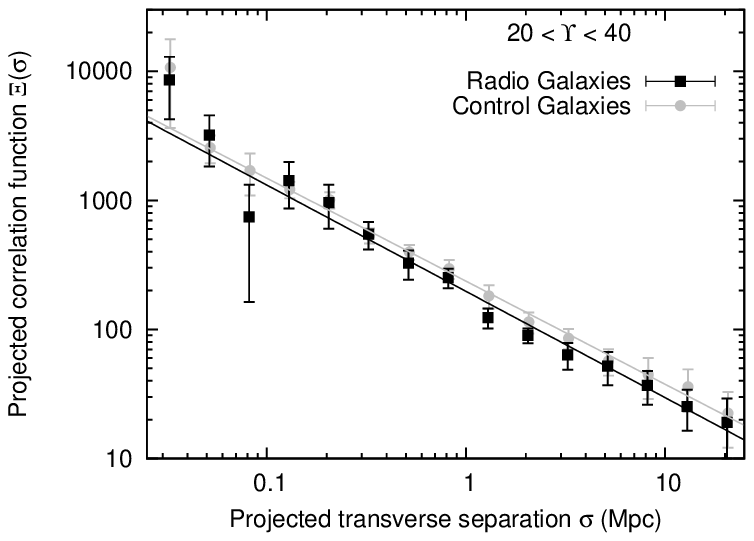}\\
\includegraphics[width=80mm]{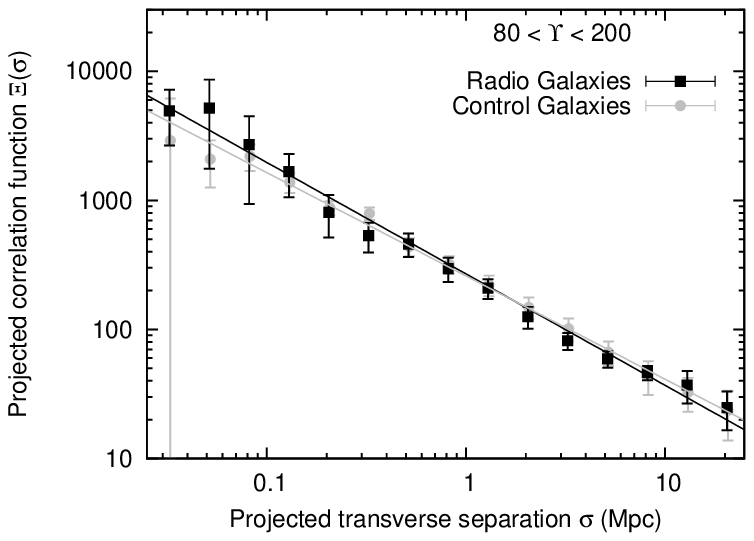}\includegraphics[width=80mm]{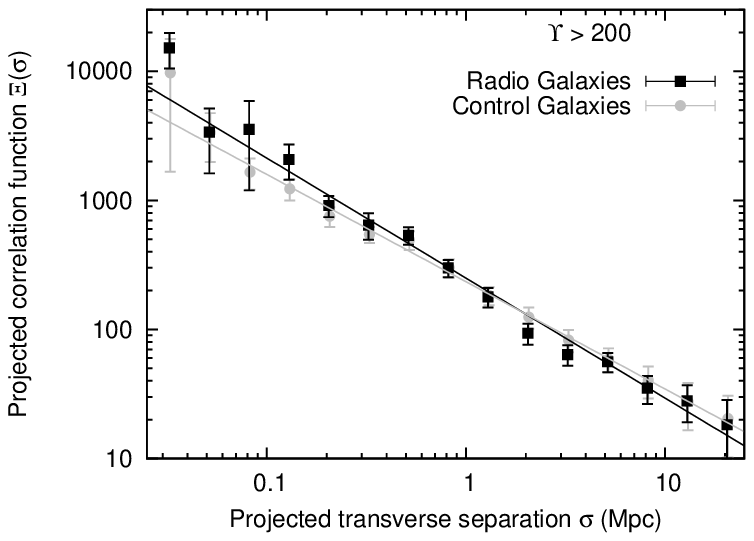}\\
\includegraphics[width=80mm]{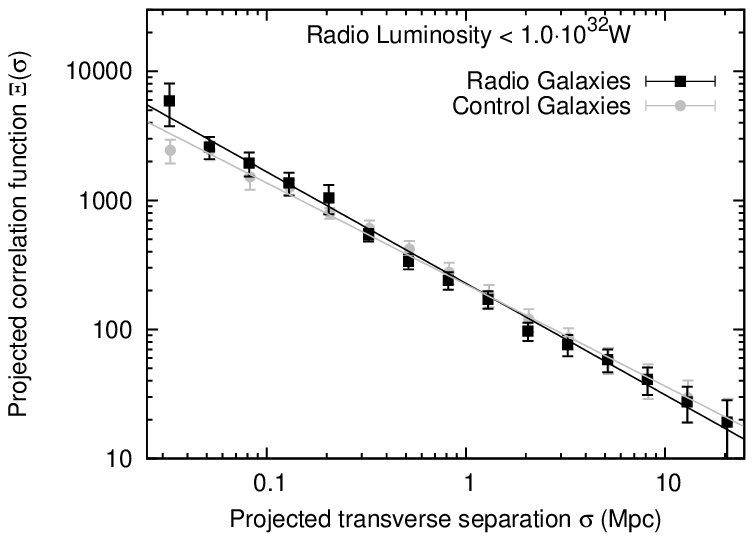}\includegraphics[width=80mm]{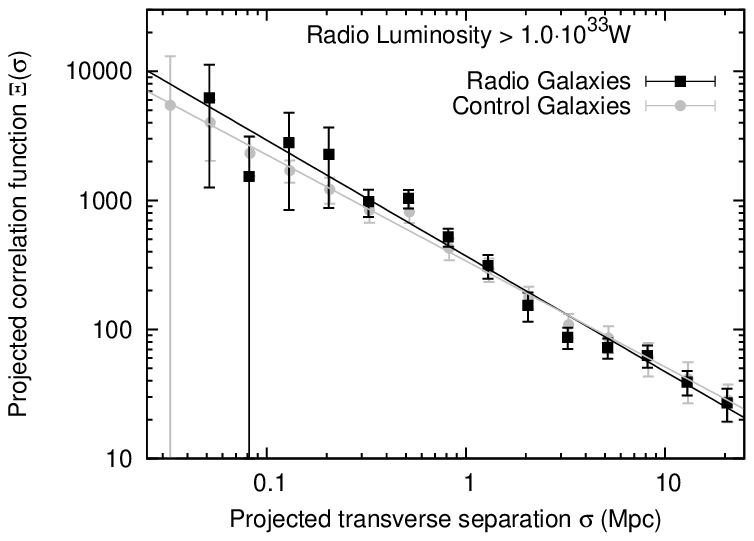}\\
\caption{ The projected correlation function $\Xi(\sigma)$ as a function of $\sigma$ for radio AGN hosts and their associated controls at four different $\Upsilon$ observed-to-median radio luminosity ratio bins. For low relative radio power the clustering properties of the two populations are virtually indistinguishable. As the relative radio luminosity increases the clustering at small scales appears to increase over the controls, though the difference does not seem statistically significant. Power law fits to the fourteen data points between 50~kpc and 22~Mpc are also shown. Also included are plots of $\Xi(\sigma)$ as a function of $\sigma$ at two different radio luminosity restrictions. There is no significant difference between the radio and control populations at either luminosity restriction.}
\label{fig:Xis}
\end{figure*}

Projected correlation functions $\Xi(\sigma)$ for radio galaxies and their associated controls at four $\Upsilon$ bins and two radio luminosity bins are shown in Figure \ref{fig:Xis}. For low and moderate radio power ($\Upsilon<200$) there is no evidence of any difference between the populations. Only for the highest relative radio power we begin to see an apparent excess at low radii for the radio AGN hosts (within about 160~kpc). For $\Upsilon>200$, the clustering strength around radio galaxies for radii between 100~kpc and 160~kpc appears greater than that of the controls, but this is not statistically significant. Beyond this distance there is also no significant difference between the two populations for any $\Upsilon$ bin. This may suggest enhanced clustering at close ranges for the most powerful radio galaxies but are not statistically significant (especially since adjacent $\sigma$ bins are somewhat correlated with each other).

\begin{figure*}
\includegraphics[width=90mm]{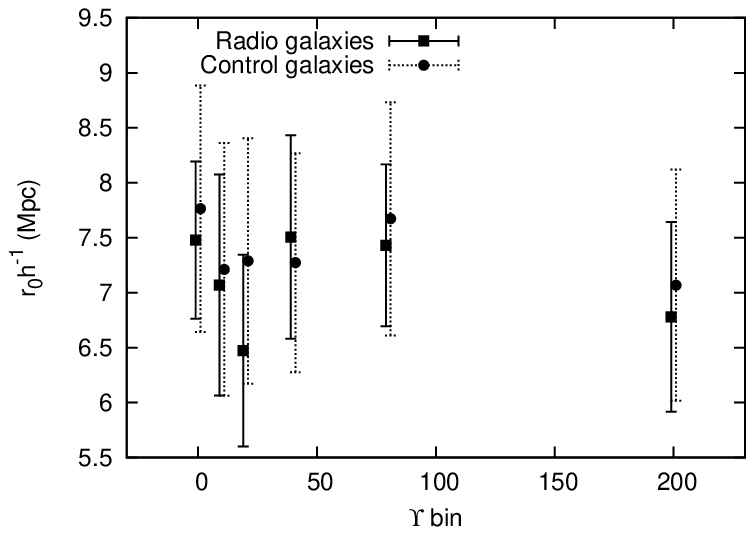}
\includegraphics[width=90mm]{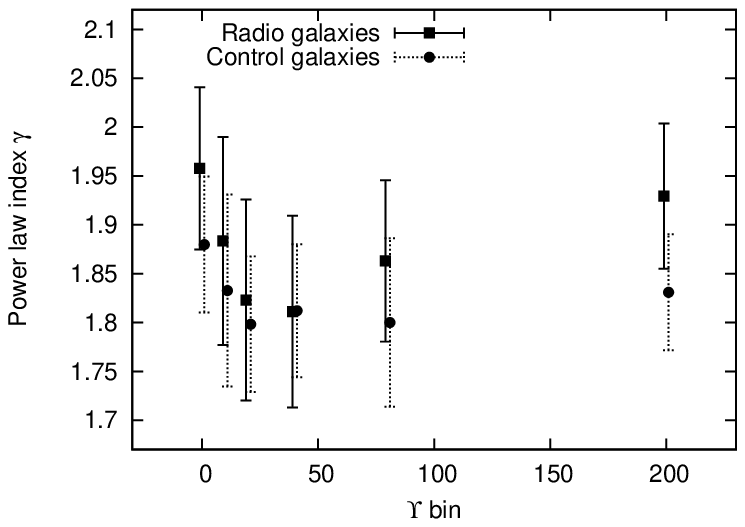}\\
\caption{ $r_0$ and $\gamma$ for the radio population at varying $\Upsilon$, and their associated controls. There is no evidence of any difference in $r_0$ between the two populations, or that $r_0$ changes with increasing $\Upsilon$. The $\gamma$ values for the RLAGN population no different from the controls, within formal errors.}
\label{fig:r0gammas}
\end{figure*}

The $r_0$ and $\gamma$ values are shown in Figure \ref{fig:r0gammas} and Table \ref{tab:r0gammas}. Reduced $\chi^2$ statistics for the power law fits to $\Xi(\sigma)$ are also given in Table \ref{tab:r0gammas}. We find no evidence of any difference in $r_0$ between the two populations, or any variation with increasing $\Upsilon$. Within formal uncertainties, the $r_0$ values for radio and control galaxies are consistent with being constant ($7.1\pm0.8$ and $7.3\pm0.9$ kpc respectively, with reduced $\chi^2$ statistics of 0.3 and 0.1 respectively for the best fit to a constant). The slightly high reduced $\chi^2$ values for the radio galaxies suggests that a single power law may not be the best fit to their correlation functions. The presence and strength of radio AGN activity are not strongly dependent on the mass of the halo in which the galaxy resides. We cannot rule out differences in $r_0$ at the 20\% level. The $\gamma$ power law indices are also indistinguishable. For the radio galaxies we get $\gamma=1.89\pm0.08$ and for the control galaxies $\gamma=1.83\pm0.06$ (with reduced $\chi^2$ of 0.4 and 0.2 respectively). The measured $\gamma$ values are probably slightly lower than the true ones because we have not corrected for fiber collisions. The difference in $\gamma$ is greatest for the $\Upsilon>200$ bin, again hinting at increased clustering at low radii for the most powerful $\Upsilon$ subset, but the errors are such that no statistically robust conclusion can be drawn.

\begin{deluxetable}{r|rrr|rrr}
%this table generated: Tue Apr 9 17:27:13 EST 2013
%\tabletypesize{\scriptsize}
\tablecolumns{7}

\tablecaption{Summary of $r_0$ and $\gamma$ values calculated from power law fits to $\Xi(\sigma)$}
\tablehead{
\colhead{ Restriction\tablenotemark{a}} & 
\colhead{Radio} & 
\colhead{Radio} & 
\colhead{Radio} \vline & 
\colhead{Control} & 
\colhead{Control} & 
\colhead{Control}  
\\
&
\colhead{$r_0$}($h^{-1}$~Mpc) &
\colhead{$\gamma$} &
\colhead{$\chi^2_\nu$}\vline &
\colhead{$r_0$}($h^{-1}$~Mpc) &
\colhead{$\gamma$} &
\colhead{$\chi^2_\nu$}
  }  
\startdata               
$\Upsilon$~<~10 & 7.48$\pm$0.71 & 1.96$\pm$0.08 & 1.68 & 7.76$\pm$0.71 & 1.88$\pm$0.08 & 1.85 \\
10~<~$\Upsilon$~<~20 & 7.07$\pm$1.00 & 1.88$\pm$0.11 & 1.16 & 7.21$\pm$1.00 & 1.83$\pm$0.11 & 0.47 \\
20~<~$\Upsilon$~<~40 & 6.47$\pm$0.87 & 1.82$\pm$0.10 & 1.47 & 7.29$\pm$0.87 & 1.80$\pm$0.10 & 0.73 \\
40~<~$\Upsilon$~<~80 & 7.51$\pm$0.93 & 1.81$\pm$0.10 & 1.56 & 7.27$\pm$0.93 & 1.81$\pm$0.10 & 0.33 \\
80~<~$\Upsilon$~<~200 & 7.43$\pm$0.74 & 1.86$\pm$0.08 & 1.18 & 7.67$\pm$0.74 & 1.80$\pm$0.08 & 0.68 \\
200~<~$\Upsilon$& 6.78$\pm$0.86 & 1.93$\pm$0.07 & 1.87 & 7.07$\pm$0.86 & 1.83$\pm$0.07 & 0.75 \\
L<$1.0\cdot 10^{32}$W& 6.78$\pm$0.85 & 1.86$\pm$0.06 & 1.98 & 7.11$\pm$.61 & 1.79$\pm$0.06 & 0.71 \\
L>$1.0\cdot 10^{33}$W& 8.57$\pm$0.80 & 1.90$\pm$0.12 & 4.26 & 8.73$\pm$0.80 & 1.82$\pm$0.12 & 0.91 \\
\hline
\enddata
\tablenotetext{a}{$\Upsilon$ is defined (see equation \ref{eqn:PhiDefinition}) as the ratio of the galaxy's observed radio luminosity to the median radio luminosity of galaxies of the same K-band magnitude.}
\label{tab:r0gammas}
\end{deluxetable}

\begin{figure}
\includegraphics[width=90mm]{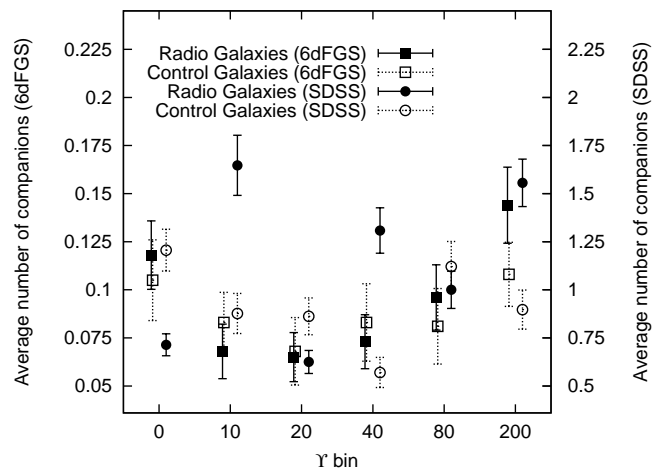}
\caption{Mean numbers of companions to radio and control galaxies from the 6dFGS and SDSS coverage. Errors are assumed to be normal and proportional to the inverse square root of the companion counts. The mean number of companions around radio galaxies is roughly the same as for controls for 6dFGS galaxies, with a marginally significant difference only for $\Upsilon>200$. There are about ten times more companions per radio/control galaxy in the SDSS data than in 6dFGS, indicating that small companions are much more common than galaxies of similar stellar mass to the radio/controls. See also Table 1.}
\label{fig:companions}
\end{figure}

To investigate the numbers of close companions, we counted the number of radio galaxy-main sample and control-main sample pairs for 10~kpc~<~$\sigma$~<160~kpc and $\pi$~<~12~Mpc in all $\Upsilon$ bins, 160~kpc being the radius of one of the $\sigma$ bins. The results are given in Table \ref{tab:CompanionCounts} and Figure \ref{fig:companions}. For a $\Upsilon$ bin of less than 200, there is no significant difference in the number of companions within 160~kpc. As the radio power increases this behaviour changes. The percentage of radio AGNs possessing a close companion increases to 14.4\% for $\Upsilon>200$ while the percentage for the controls climbs more slowly, to 10.8\%. For $\Upsilon>200$, one in 27 RLAGNs has an extra close companion compared to the associated controls. These results are stronger ($2.2\sigma$) than the correlation function measure. For the absolute radio power selected bins, we found no significant difference between the radio and control populations. If we count companions within 100 or 50~kpc of the radio/control galaxy the results are similar but with reduced significance.

We emphasize here that it is not the absolute numbers of detected companions that matters, but the relative \emph{difference} in companion counts between the controls and radio galaxies. Our $\Upsilon$ bins differ in median redshift (see Table \ref{tab:CompanionCounts}), so our ability to detect companions of a given stellar mass also differs between $\Upsilon$ bins. The absolute numbers are affected by fiber collisions, as discussed in \S \ref{sec:CrossCor}, and the range at which fiber collisions affect the numbers is also redshift dependent. However, since the radio and control populations are well-matched in redshift and radio galaxies are not selected differently to the main sample in 6dFGS, these issues affect both populations equally.

The mean difference in K-band magnitude between the radio galaxies and their companions is 0.17 for $\Upsilon>200$, indicating that the companions are of comparable mass to the RLAGN hosts. This is to be expected, since the 12.9 magnitude K-band restriction on the main sample galaxies is bright. Our analysis is therefore not sensitive to fainter companions.

To quantify the effect of excluding faint companions we looked for objects near the RLAGN hosts in question in the Sloan Digital Sky Survey, Data Release 7 \citep{AbazajianEtAl2009}. This is primarily a northern survey but has some overlap with 6dFGS; it has spectroscopic data for $\sim$~55,000 southern galaxies nearer than our main sample limit of $z=0.27$. The SDSS data extends to galaxies over two magnitudes fainter in the R band than 6dFGS. We restricted the radio and control samples to the densest areas of the SDSS coverage. We performed pair counts as before, for all the $\Upsilon$ bins. Visual inspection of SDSS images also gives us confidence that the level of contamination is not significant.

Figure \ref{fig:GalaxyPics} displays a selection of four RLAGNs and four controls for $\Upsilon>200$, chosen from our catalogs in order of decreasing declination. The increased numbers of companions around the highest $\Upsilon$ galaxies, compared with controls, are clearly visible.

As seen in Table \ref{tab:CompanionCounts} and Figure \ref{fig:companions}, there are about ten times as many faint SDSS companions as bright 6dFGS companions per radio/control galaxy. The limited area of overlap between SDSS and 6dFGS means that the absolute companion counts are small and there is considerable scatter in the data. The most significant results for the relative radio luminosity subsets are in the $\Upsilon<10$ bin, which shows less companions than the controls at $3.1\sigma$ significance, and the $\Upsilon>200$ bin, which shows almost twice as many companions for the radio galaxies at $2.2\sigma$ significance. However, due to the small numbers of galaxies in the 6dFGS-SDSS overlap area and large scatter, we cannot draw any strong conclusions from these results.

\begin{deluxetable*}{rrrrr|rrrr}
%this table generated: Tue Apr 9 17:33:52 EST 2013
\rotate{270}
\tabletypesize{\scriptsize}
\tablecolumns{9}

\tablecaption{Number of companions within a projected distance of 160~kpc to radio galaxies and their associated controls per $\Upsilon$ bin}
\tablehead{
  \colhead{Restriction}                   &
  \colhead{Radio} &
  \colhead{Companions} &
  \colhead{Median z} &
  \colhead{Median $M_K$}&
 \colhead{Control}      &
  \colhead{Companions}   &
  \colhead{Median z} &
  \colhead{Significance\tablenotemark{a}}
  \\
  &
  \colhead{galaxies} &
  &
  &
  &
  \colhead{galaxies}      &
  &
  &
  }  
\startdata               
$\Upsilon$~<~10 &373 &  44 (11.8\%) &  0.095 &-25.96  & 1492 &  157 (10.5\%) & 0.094299 & 0.6\\
10~<~$\Upsilon$~<~20 &336 &  23 (6.8\%) &  0.089 &-25.57  & 1344 &  111 (8.3\%) & 0.088244 & 1.0\\
20~<~$\Upsilon$~<~40 &399 &  26 (6.5\%) &  0.0832 &-25.36  & 1596 &  109 (6.8\%) & 0.083698 & 0.2\\
40~<~$\Upsilon$~<~80 &371 &  27 (7.3\%) &  0.080 &-25.32  & 1484 &  123 (8.3\%) & 0.080282 & 0.6\\
80~<~$\Upsilon$~<~200 &332 &  32 (9.6\%) &  0.077 &-25.27  & 1328 &  107 (8.1\%) & 0.077427 & 0.8\\
$\Upsilon$~>~200 &367 &  53 (14.4\%) &  0.073 &-25.14  & 1468 &  159 (10.8\%) & 0.072687 & 1.6\\
L<$1.0\cdot 10^{32}$W&963 &  130 (13.5\%) &  0.064 &-25.13  & 3872 &  522 (13.5\%) & 0.063484 & 0.0\\
L>$1.0\cdot 10^{33}$W&353 &  17 (4.8\%) &  0.108 &-25.93  & 1412 &  54 (3.8\%) & 0.107147 & 0.7\\
\hline
\multicolumn{5}{c}{Companions in}\\
\multicolumn{5}{c}{SDSS DR7\tablenotemark{b}}\\
\hline
$\Upsilon$~<~10 &35 &  25 (71.4\%) &  0.104 &-26.16  & 102 &  123 (120.6\%) & 0.082534 & 3.1\\
10~<~$\Upsilon$~<~20 &17 &  28 (164.7\%) &  0.089 &-25.73  & 81 &  71 (87.7\%) & 0.080055 & 2.1\\
20~<~$\Upsilon$~<~40 &24 &  15 (62.5\%) &  0.083 &-25.32  & 94 &  81 (86.2\%) & 0.083968 & 1.3\\
40~<~$\Upsilon$~<~80 &13 &  17 (130.8\%) &  0.084 &-25.45  & 93 &  53 (57.0\%) & 0.080542 & 2.0\\
80~<~$\Upsilon$~<~200 &17 &  17 (100.0\%) &  0.073 &-25.48  & 66 &  74 (112.1\%) & 0.075345 & 0.4\\
$\Upsilon$~>~200 &27 &  42 (155.6\%) &  0.079 &-25.45  & 87 &  78 (89.7\%) & 0.081386 & 2.2\\
L<$1.0\cdot 10^{32}$W&58 &  55 (94.8\%) &  0.078 &-25.26  & 235 &  302 (128.5\%) & 0.065299 & 2.4\\
L>$1.0\cdot 10^{33}$W&25 &  31 (124.0\%) &  0.086 &-26.11  & 88 &  74 (84.1\%) & 0.100973 & 1.5\\
\hline
\enddata
\tablenotetext{a}{The significance of the difference in companion counts between radio and control populations. Errors are assumed to be proportional to the inverse square root of the companion count.}
\tablenotetext{b}{The reduced numbers of galaxies in the radio and control populations is due to the small area of overlap between the 6dFGS and SDSS surveys.\\[8mm]}
\label{tab:CompanionCounts}
\end{deluxetable*}

\section{Discussion}
\label{sec:Discussion}

\begin{figure*}
\centering
{\includegraphics[width=0.22\textheight,angle=0]{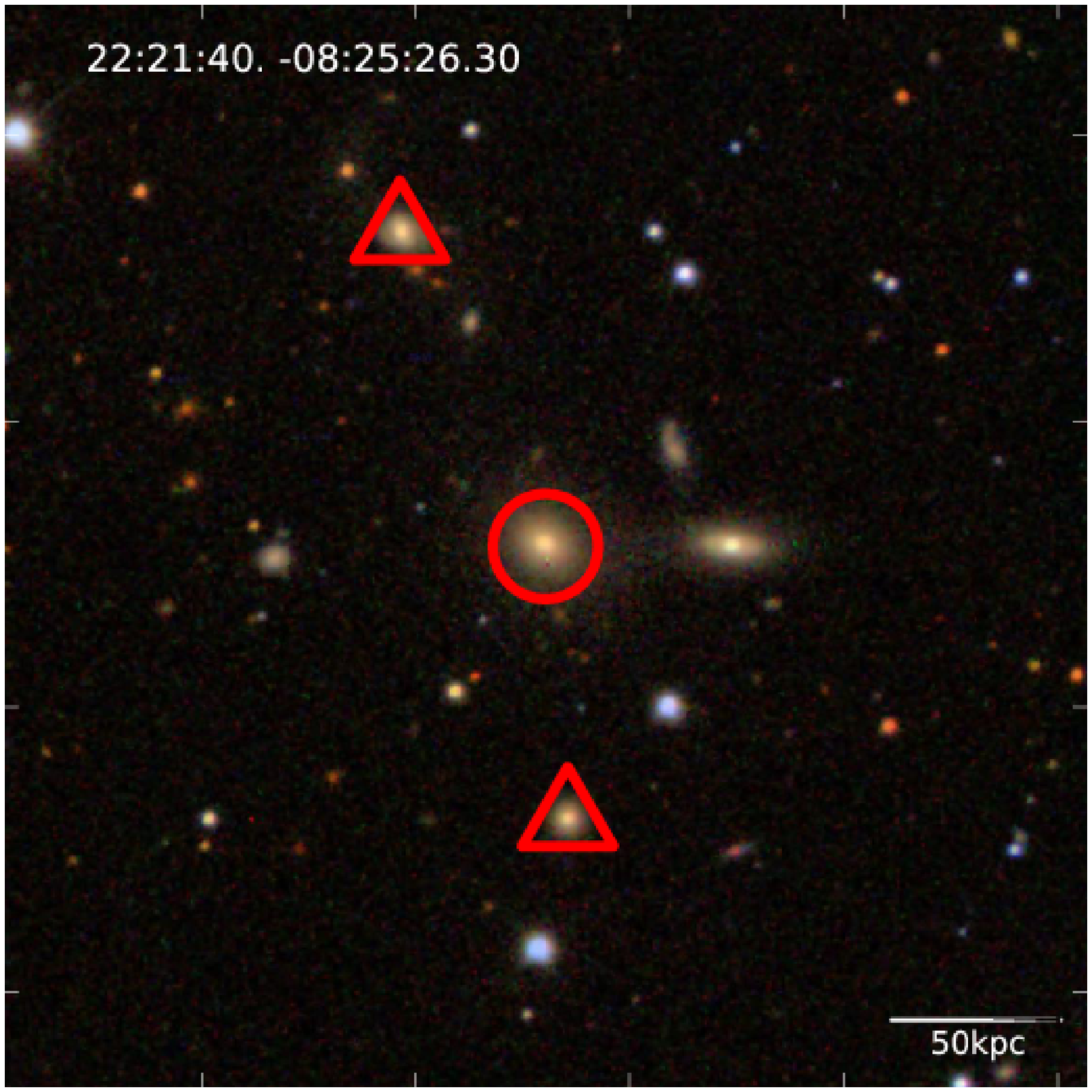}}{\includegraphics[width=0.22\textheight,angle=0]{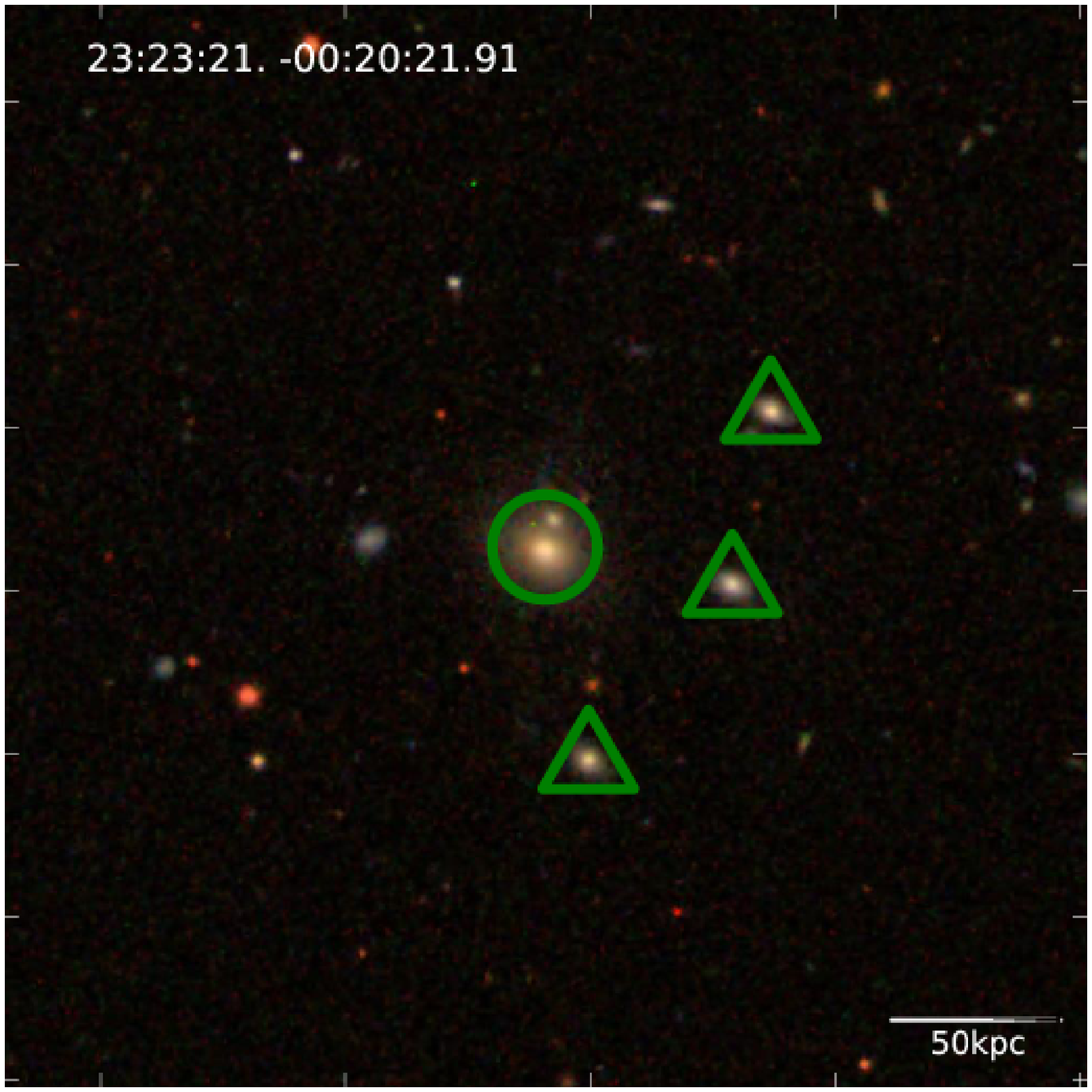}}\\
{\includegraphics[width=0.22\textheight,angle=0]{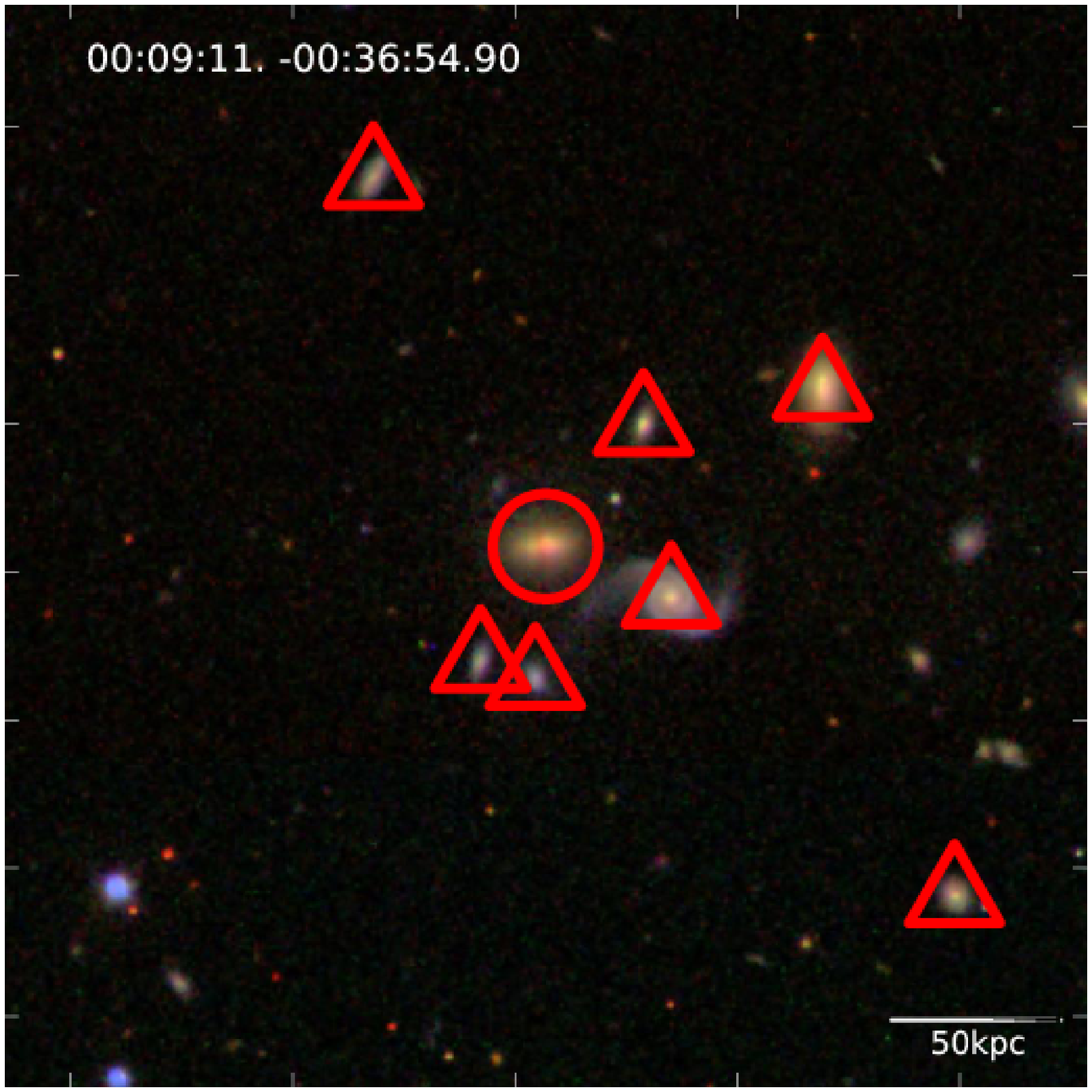}}{\includegraphics[width=0.22\textheight,angle=0]{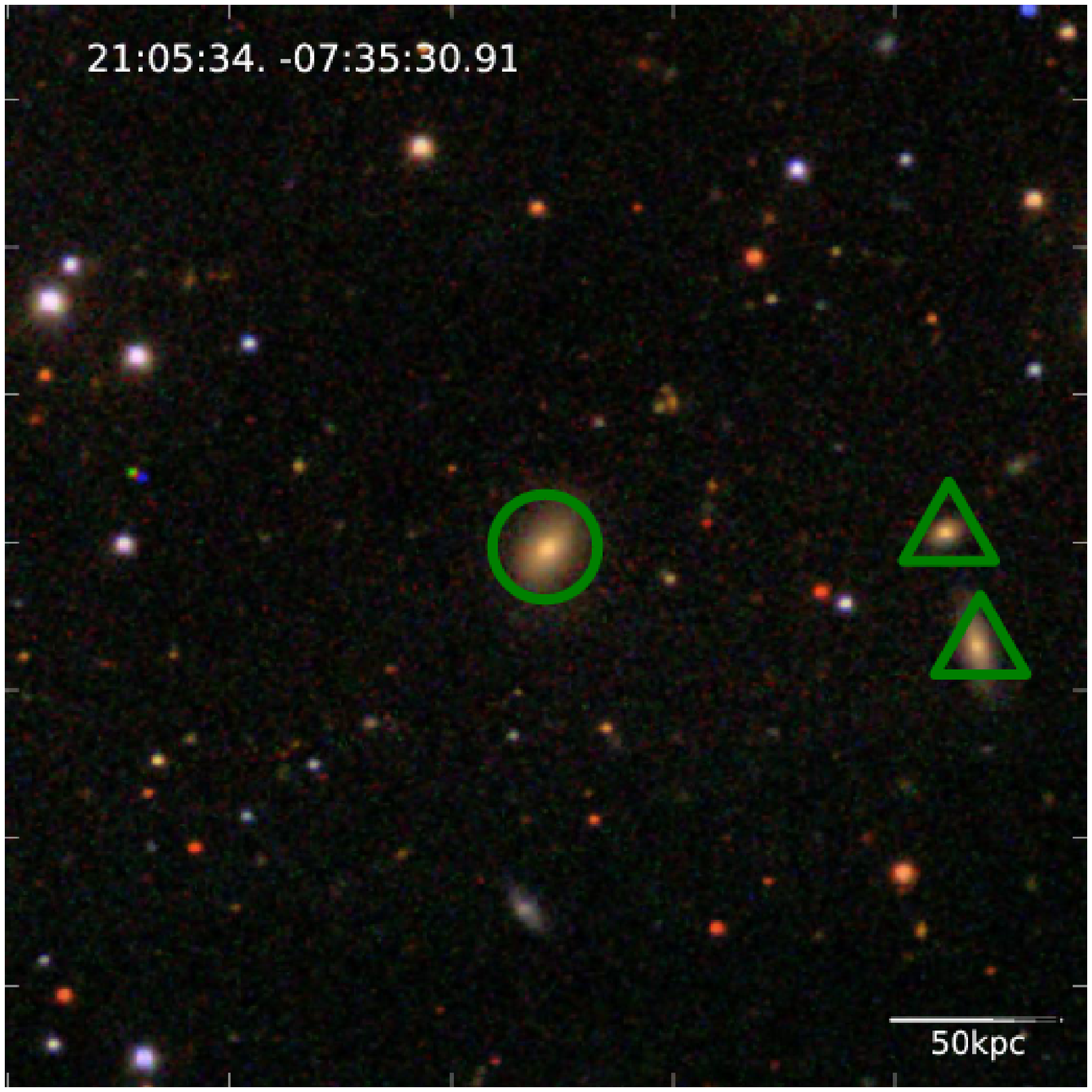}}\\
{\includegraphics[width=0.22\textheight,angle=0]{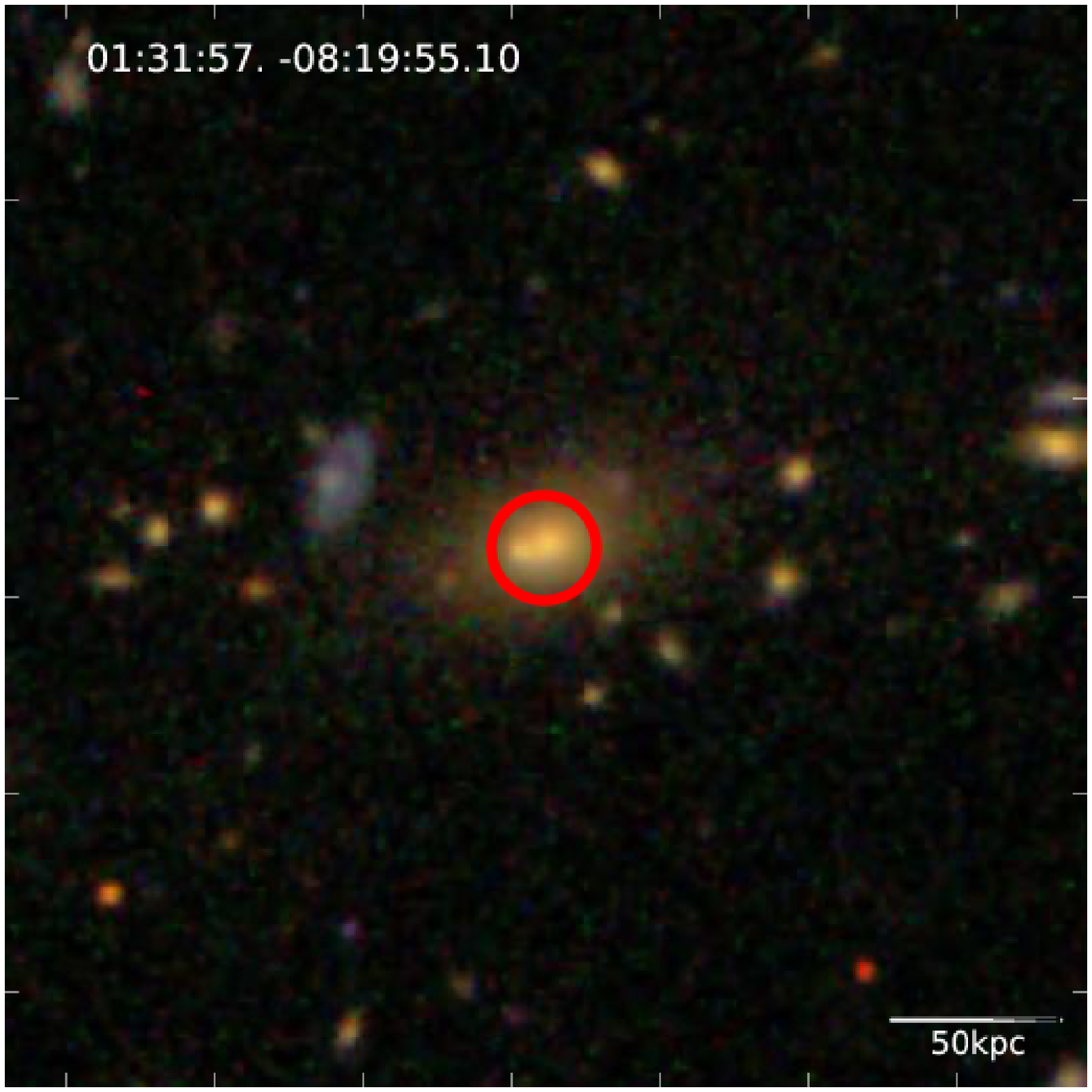}}{\includegraphics[width=0.22\textheight,angle=0]{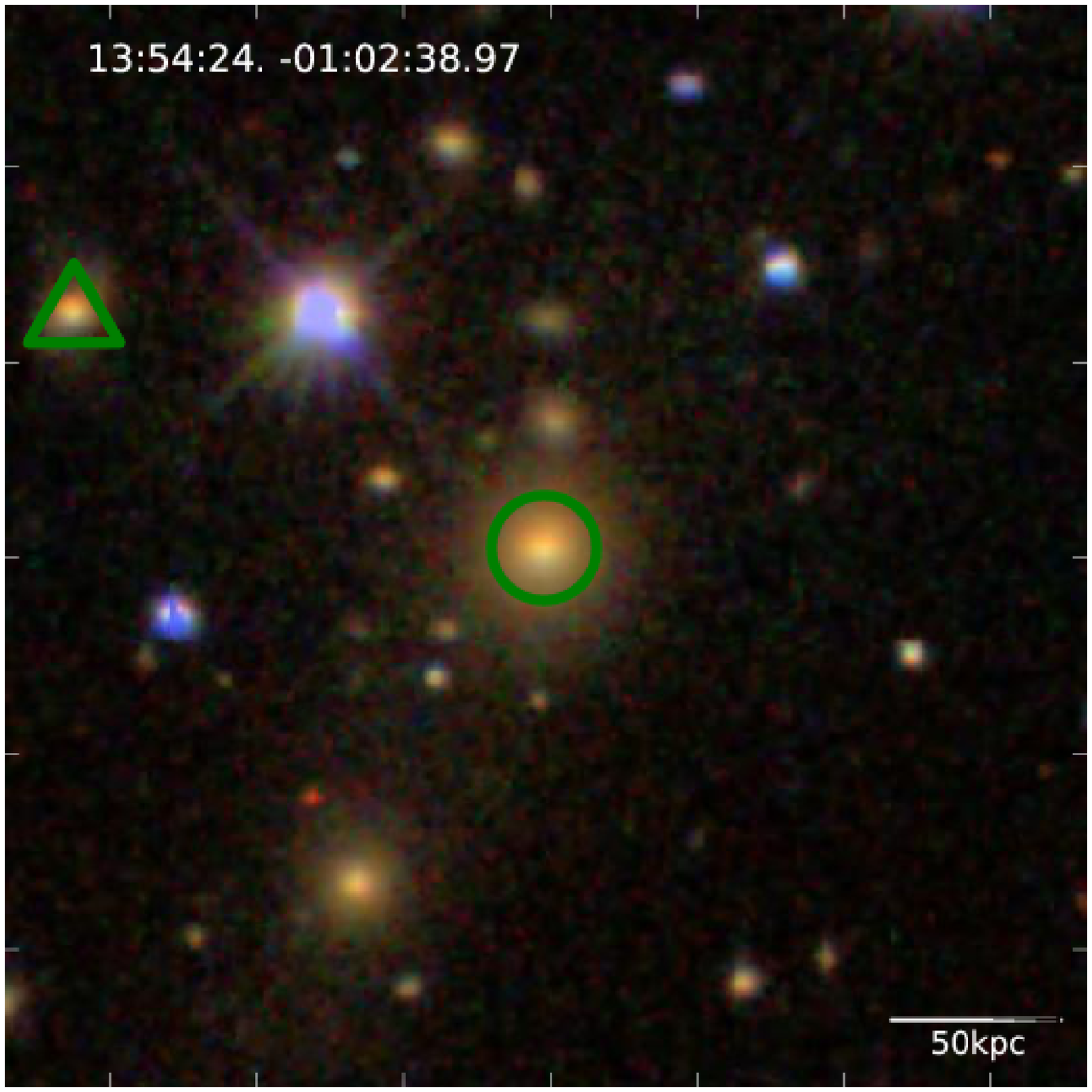}}\\
{\includegraphics[width=0.22\textheight,angle=0]{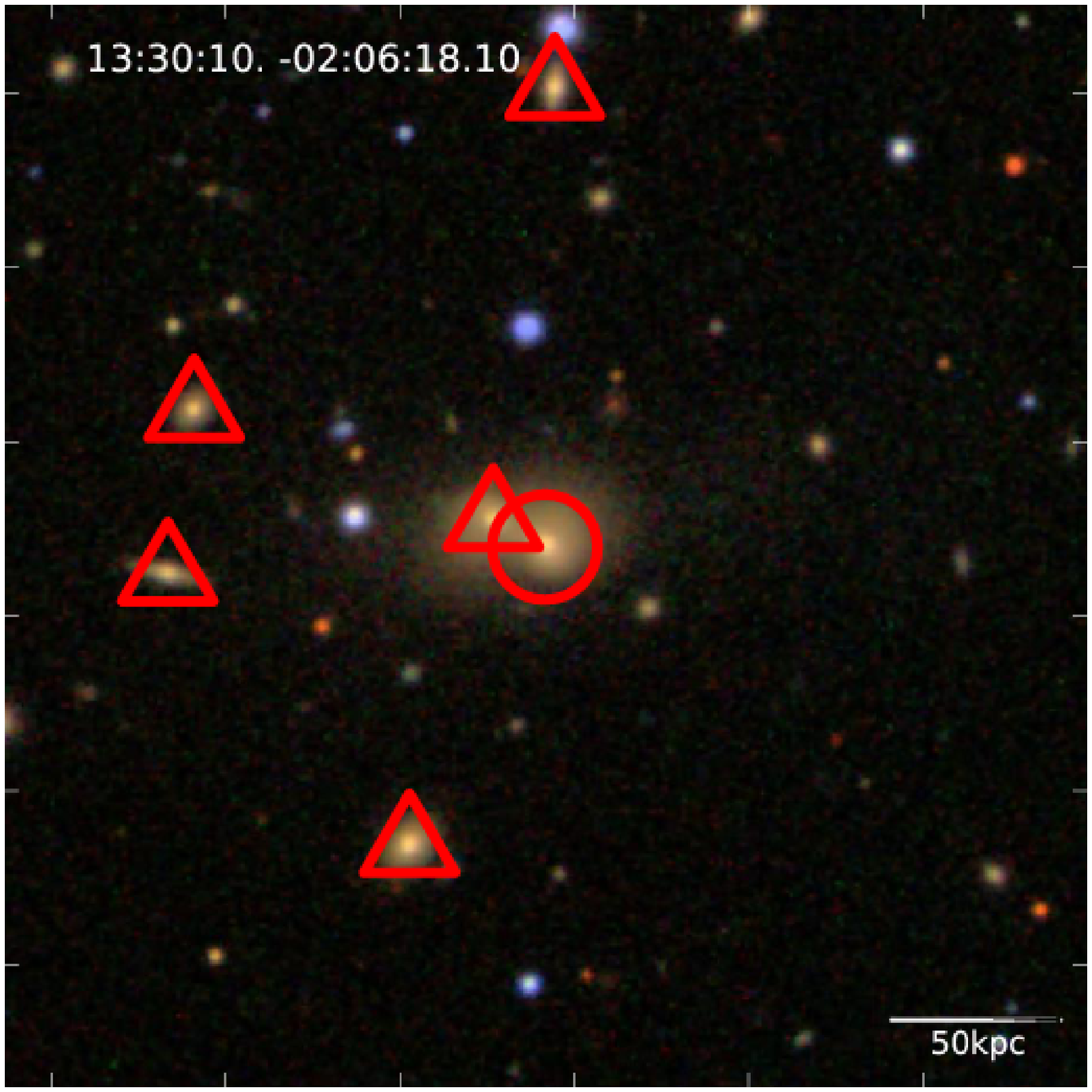}}{\includegraphics[width=0.22\textheight,angle=0]{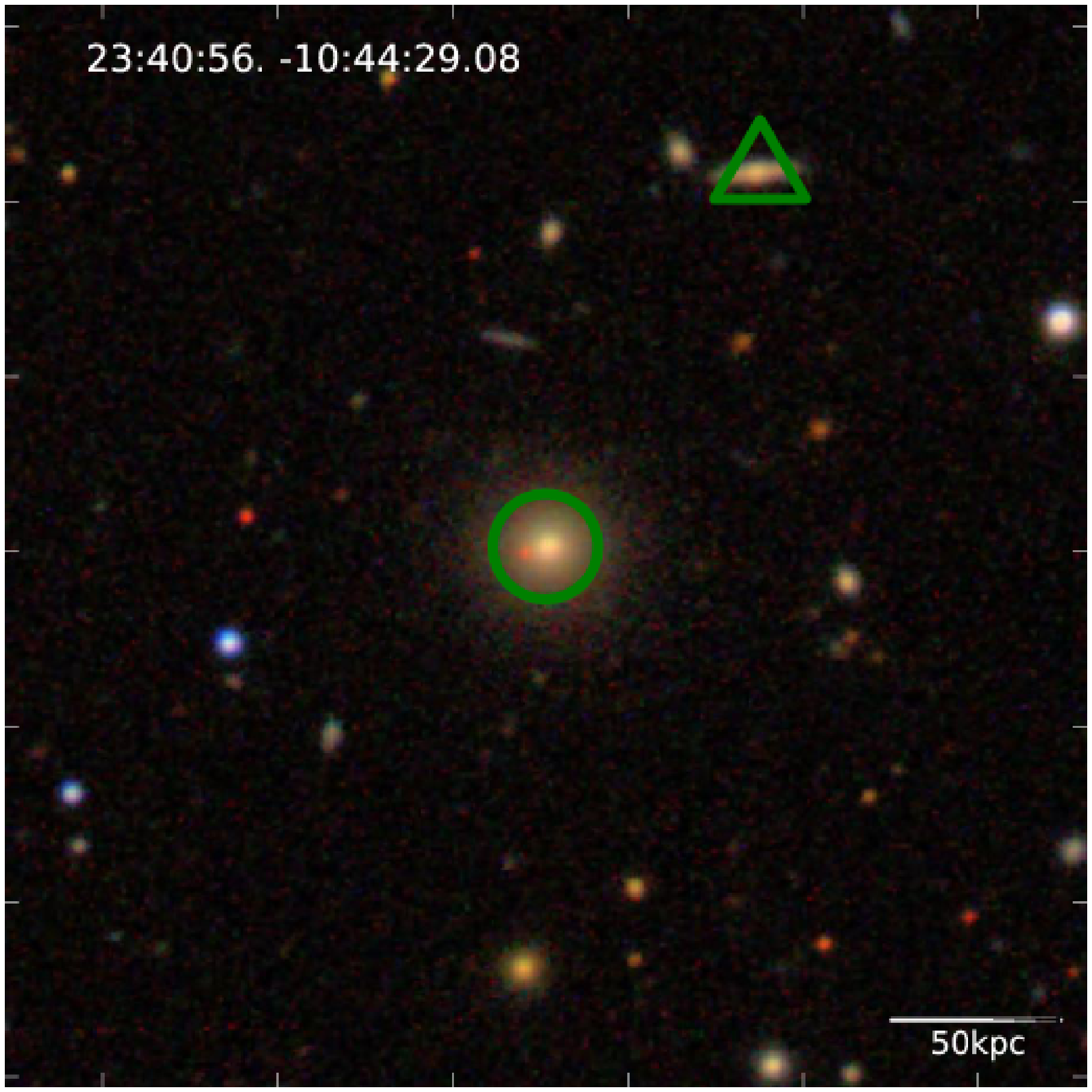}}

\caption{Selected SDSS images of RLAGN hosts (left columns) and their associated controls (right columns) for $\Upsilon>200$. All images are scaled to 320~kpc a side. The central galaxies are marked with circles and SDSS companions with triangles. It appears that the radio galaxies generally have more close companions than the controls, though the counts may be dominated by a few radio galaxies with many companions. There are radio galaxies with no companions and control galaxies with several. Although there are no disk galaxies among the radio or control galaxies in this figure, a small level of contamination is still likely present. Visual inspection of images like this was critical in verifying that only valid companion galaxies were included, and that disk galaxy contamination was low. SDSS spectral data is incomplete, meaning that some visible companions are not marked.}
\label{fig:GalaxyPics}
\end{figure*}

We have found no significant environmental dependence for $\Upsilon<200$ galaxies. Radio emission does not appear to depend significantly on either halo mass or the presence of nearby galaxies. We consider whether secular evolution is taking place, in a scenario where ellipticals wander somewhat randomly in radio power over many Gyr (e.g., \citealt{BestEtAl2007,ShabalaEtAl2008}). Only for $\Upsilon>200$ do we see some evidence for an environmental dependence (on <160 kpc scales only) consistent with ongoing mergers boosting radio power. On average, RLAGNs with $\Upsilon>200$ have about 33\% more 6dFGS companions than the control population, and almost twice as many SDSS companions. This applies to the overall population of radio sources, but there are individual radio sources with $\Upsilon>200$ that do not have nearby companions (see Fig \ref{fig:GalaxyPics}) and controls that have many nearby companions. Galaxies binned by their radio luminosity exhibit no significant differences from their associated controls. This confirms the importance of accounting for their relative, rather than their absolute, radio powers. One consequence of defining the sample as we have done is seen in figure \ref{fig:companions}, where the companion numbers seem to decrease, then increase, with increasing $\Upsilon$. For the lowest $\Upsilon$ values we are selecting galaxies that are radio loud in an absolute sense (>2.8~mJy) but radio quiet in a relative sense, i.e. they must be large galaxies and therefore have more companions simply because of their size.  Table \ref{tab:CompanionCounts} shows that the $\Upsilon<10$ subset is, on average, about twice as luminous as the $\Upsilon>200$ subset. This effect diminishes at higher $\Upsilon$.

We have found no evidence of a difference in clustering strength beyond about 160~kpc for any $\Upsilon$ or absolute radio luminosity subset. This is roughly the scale where the correlation function is dominated by the relationship between host dark matter halo mass and large-scale bias rather than the arrangement of galaxies within individual haloes \citep{ZehaviEtAl2004}. This result for the large-scale clustering implies that radio galaxies have host dark matter halo masses comparable to those of the control galaxies selected with comparable colors, absolute magnitudes, and (approximate) morphology. We find that if reddened edge-on spiral galaxies (which are easy to overlook) are not removed, one can get a spurious difference between radio galaxies and control galaxies.

For a given galaxy stellar mass, we would expect the host halo masses of central and satellite galaxies to differ, so our result appears to conflict with the idea that radio galaxies are preferentially central galaxies (e.g., \citealt{BestEtAl2007}). However, the median absolute magnitude of our sample is always brighter than $M_K=-25$, and HOD modelling indicates our radio galaxies and controls are dominated by central galaxies \citep{ZhengEtAl2009}. \cite{BestEtAl2007} also find that the fractions of central galaxies and all galaxies that host a radio AGN are comparable as one moves to high mass galaxies. Thus, our results actually agree with with those of these authors.

Since we found similar $r_0$ values for all our galaxy subsets (and both galaxies in the radio-galaxy and control-galaxy pairs have similar stellar masses) we can meaningfully compare our results even with auto-correlation studies performed without regard to radio emission. \cite{ZehaviEtAl2011} considered the auto-correlation of a wide variety of galaxy types in the SDSS catalog. Their ''redder'', ''redseq'', and ''reddest'' subsets (see their Figure 18 and Table 4) are physically similar to both radio and control samples in our catalogues. They find $r_0$ ranging between 5.48~$h^{-1}$~Mpc (for the ''redder'' set) and 7.62~$h^{-1}$~Mpc ( for the ''reddest'' set). These $r_0$ values are similar to ours. \cite{ZehaviEtAl2011} found slightly higher $\gamma$ than we did, ranging from 1.9 to 2.1 for the ''redder'' and ''reddest'' subsets respectively. Since we have not corrected for fiber collisions, our $\gamma$ values are probably artificially low. Our results are therefore in  broad agreement with those of \cite{ZehaviEtAl2011}.

\cite{WakeEtAl2008} and \cite{DonosoEtAl2010} find that RLAGNs cluster more strongly than controls at all scales, implying that radio-loud AGNs occupy more massive dark matter haloes than radio-quiet controls and contrasting with our results. Both of these studies also find that the difference between RLAGNs and controls is most pronounced within about 1~Mpc. The small sample size of \cite{WakeEtAl2008} prevented that study from detecting differences in the projected correlation function between radio and control samples of better than about $2\sigma$, or to investigate clustering at scales of less than about 500~kpc. \cite{HickoxEtAl2009} was similarly hampered by a small RLAGN sample size; they could not probe close-range clustering, and found no difference in large scale clustering. They found $r_0$ values for radio AGNs of $6.3\pm0.6h^{-1}$, which are similar to ours. A closer inspection of the \cite{DonosoEtAl2010} paper (their Figure 3) shows that the projected correlation function for radio galaxies exceeds that for the controls by about 25\% at distances greater than 1~Mpc, whereas our study cannot rule out differences of up to about 15\%. They also found that the projected correlation function within 1~Mpc was around 75\% greater for the radio galaxies than the controls, which is of the same order as our result of $\sim$~40\%. We therefore conclude that the discrepancies between the \cite{WakeEtAl2008} and \cite{DonosoEtAl2010} studies and ours are not very significant. Importantly, \cite{DonosoEtAl2010} do not exclude disk contaminants from their control samples, which may exaggerate the difference between these two populations. Both \cite{DonosoEtAl2010} and \cite{WakeEtAl2008} study galaxies at $z\approx 0.55$ and radio luminosities in excess of $10^{24}$~W~Hz$^{-1}$, that is, significantly more distant and brighter than the galaxies in our study. Characterising the transition between these two regimes is clearly of interest. For convenience we provide a summary of these studies in Table \ref{tab:StudyComparison}.

\begin{deluxetable*}{rrrrrrrr}
\tabletypesize{\scriptsize}
\tablecolumns{8}

\tablecaption{Comparison of RLAGN galaxy clustering studies}
\tablehead{
  \colhead{Study}                   &
  \colhead{Survey(s)} &
  \colhead{No. of RLAGN} &
  \colhead{Redshift $z$} &
  \colhead{Radio}&
  \colhead{Visual}      &
  \colhead{$r_0$}   &
  \colhead{$\gamma$} \\
  &
  &
  &  
  & \colhead{Lum. (W~Hz$^{-1}$)}
  & \colhead{Magnitude}
  & \colhead{(h$^{-1}$ Mpc)}
  &
  }  
\startdata               
This Study & 1,2,3 &2,178 & 0.02-0.27 & $10^{21.3}-10^{25.8}$ & $m_K$ < 12.9&$7.1\pm0.9$ & $1.89\pm0.08$ \\
\cite{DonosoEtAl2010} & 3 & 14,453 & 0.4-0.8 & > $10^{24}$ & $m_I$ < 20 & $8.4\pm0.3$\tablenotemark{a}&$1.81\pm0.05$\\
\cite{WakeEtAl2008} & 4 & 250 & 0.4-0.8 & $10^{23}-10^{26}$ & $m_I$ < 19.8& $9.6\pm 0.5$ & $1.75\pm 0.10$ \\
\cite{HickoxEtAl2009} & 5,6 & 122 & 0.25-0.8 & $10^{23.8}-10^{26.3}$ & $m_I$ < 20 & $6.3\pm0.6$ & $1.8\pm0.2$ \\
\cite{MagliocchettiEtAl2004} & 7,8 & 536 & 0.01-0.3 & $10^{20}-10^{25}$& $b_J$ < 19.37 & $7.6\pm 0.1$\tablenotemark{b} & $2\pm0.1$\\
\cite{MandelbaumEtAl2009} & 2,4,8 & 5,712 & 0.01-0.3 & $10^{23}-10^{26}$ & 14.5 < $r$ < 17.6 & N/A & N/A\\
\enddata
\tablenotetext{a}{Calculated here using Eqn. 10 and 11 in \cite{MeyerEtAl2007}, with $h=1$}
\tablenotetext{b}{Using their $h=0.7$}
\tablerefs{1. 6dFGS: \cite{JonesEtAl2009}, 2. NVSS: \cite{CondonEtAl1998}, 3. SDSS: \cite{AbazajianEtAl2009}, 4. 2SLAQ LRG: \cite{CannonEtAl2006, SadlerEtAl2007}, 5. AGES: \cite{KochanekEtAl2012}, 6. WSRT: \cite{deVriesEtAl2002}, 7. FIRST: \cite{BeckerEtAl1995}, 8. 2dFGRS: \cite{CollessEtAl2001}}
\label{tab:StudyComparison}
\end{deluxetable*}

Furthermore, \cite{DonosoEtAl2010} find that the difference in clustering between RLAGN and controls at close ranges is more pronounced for low-mass hosts. This agrees qualitatively with our results since, for a given radio power, low mass galaxies have higher $\Upsilon$ ratios than more massive ones this further highlights the importance of considering radio luminosity \emph{relative to the stellar mass of the host galaxy} rather than just its radio luminosity.

\cite{MandelbaumEtAl2009} find that clustering of galaxies around radio AGN is stronger on both large and small scales than the clustering of galaxies around controls. This implies radio galaxies have higher halo masses than control galaxies (see their Figure 9). However, the \cite{MandelbaumEtAl2009} AGN sample probes lower mass host galaxies than our work. When \cite{MandelbaumEtAl2009} restrict their samples to galaxies with stellar mass of $>10^{11.44}$ (comparable to $M_K<-24.6$), there is little difference between the clustering of galaxies around radio AGN and the clustering of galaxies around control galaxies at any scale. For high mass host galaxies, they do not see an excess of nearby companions, in contrast to our work and \cite{DonosoEtAl2010}. We speculate that this may be due to companion galaxies being fainter than the radio galaxies and the SDSS spectroscopic magnitude limits.

\section{Conclusions}
\label{sec:Conclusions}

We have studied the influence of environment on RLAGN activity at both large and small scales to investigate whether AGN activity depends on dark matter halo mass, and whether it can be triggered by encounters with nearby companions. We used the two-point cross-correlation function to compare the numbers of companions to radio-loud AGNs in the 6dFGS survey area with a radio-quiet control population well-matched in luminosity, colour, and redshift. As part of this analysis, we have taken great care to exclude disk galaxies from the control sample. We have introduced the quantity $\Upsilon$, the ratio of the observed radio luminosity of an AGN to the median radio luminosity of galaxies with the same stellar mass. It is a means of distinguishing powerful radio activity emitted by a small galaxy from more modest activity associated with a larger one.

There is only tentative ($1.6\sigma$) evidence for an excess of nearby companion galaxies around radio galaxies, and only for those radio galaxies where the radio power is more then 200 times the median radio power. We have further investigated this, looking at SDSS companions around brighter 6dFGS radio galaxies, but the evidence is of marginal significance ($2.2\sigma$), and there is no obvious trend with increasing relative radio power. For most of the samples studied, the clustering of galaxies around radio galaxies and controls is essentially indistinguishable at any scale.

We find that the clustering properties of radio galaxies and controls are comparable for scales greater than about 160~kpc. This suggests that that these two populations inhabit dark matter haloes of similar mass. Large scale clustering is not a function of relative radio power, but we do observe a slight excess of nearby companions around the sources emitting most strongly relative to their stellar mass. Our result is consistent with radio power varying randomly by orders of magnitude over billions of years.

\section*{Acknowledgements}

H.W. is supported by an APA postgraduate research scholarship. F.B. is supported by the Australian Government through the International Postgraduate Research Scholarship (IPRS) and by scholarships from ICRAR and the AAO. M.B. and D.F. acknowledge support from the Australian Research Council via Discovery Project grant DP110102174. M.B. acknowledges the support from the Australian Research Council via Future Fellowship grant FT100100280. 

The 6dF Galaxy Survey was funded in part by an Australian Research Council Discovery--Projects Grant (DP-0208876), administered by the Australian National University. We acknowledge the use of the HyperLeda database (http://leda.univ-lyon1.fr). The National Radio Astronomy Observatory is operated by Associated Universities Inc., under co-operative agreement with the National Science Foundation.

This publication makes use of data products from the Two Micron All Sky Survey, which is a joint project of the University of Massachusetts and the Infrared Processing and Analysis Center/California Institute of Technology, funded by the National Aeronautics and Space Administration and the National Science Foundation.

We are grateful to Tom Mauch and Elaine Sadler for the use of their updated AGN catalog, and Tim Dolley for helpful discussions and suggestions. We thank the anonymous referee, whose comments led to substantial improvements in this paper.

\bibliography{ClusteringPaper.bib}
\end{document}